%
%
%

\ifx\mnmacrosloaded\undefined \input mn\fi


\pageoffset{-0.2cm}{0pc}

%
  
%
%
%



\begintopmatter  

\title{An Updated Precision Estimate of the Hubble Constant and the Age and
Density of the Universe in the
Decaying Neutrino Theory}
\author{D. W. Sciama}
\affiliation{S.I.S.S.A., I.C.T.P., Strada Costiera 11, 34014 Trieste, Italy}
\vskip 8.5pt 
\affiliation {Nuclear and Astrophysics Laboratory, University of Oxford}
\vskip 8.5pt 
\affiliation {e-mail: sciama@sissa.it}
\shortauthor{D. W. Sciama}
\shorttitle{The Decaying Neutrino Theory}


\abstract {We here update the derivation of precise values for the Hubble
constant H$_0,$ the age $t_0$ and the density parameter $\Omega h^2$ of
the universe in the decaying neutrino theory for the ionisation of the
interstellar medium (Sciama 1990 a, 1993). Using recent measurements of
the temperature of the cosmic microwave background, of the abundances of
D, He$^4$ and Li$^7,$ and of the intergalactic hydrogen-ionising photon
flux at zero red shift, we obtain for the density parameter of the
universe ${\Omega}h^2=0. 300\pm 0. 003$.  Observed limits on H$_0$ and
$t_0$ then imply that, for a zero cosmological constant, H$_0=52. 5\pm 2.
5$ km. sec$^{-1}$ Mpc$^{-1}, t_0=12. 7\pm 0. 7$ Gyr and ${\Omega}=1. 1\pm
0. 1$. If ${\Omega}=1$ exactly, then H$_0=54.8\pm 0.3$ km. 
sec$^{-1}$Mpc$^{-1}$, and $t_0=11.96\pm 0.06$ Gyr. These precise
predictions of the decaying neutrino theory are compatible with current
observational estimates of these quantities.}


\maketitle  

\section{Introduction}

Recent developments in attempts to measure the Hubble constant H$_0$ and
the age of the universe
$t_0$ suggest that it would be worth updating the precision values for
these quantities derived
(Sciama 1990 b) in the decaying neutrino theory for the ionisation of the
interstellar medium
(Sciama 1990a, 1993). Such updating can take advantage of recent reductions
in the uncertainties
of each of the quantities which enter into this derivation. These
quantities are
$n_{\nu}/n_{\gamma},n_{\gamma}, n_{b}$ and $m_{\nu}$, where $n_{\nu}$ and
$m_{\nu}$ are the
present cosmological number density and rest mass of the decaying neutrinos, while
${\gamma}$ and b refer to cosmic microwave photons and baryons
respectively. The current
uncertainties in each of the first three quantities enter into the derived
value of ${\Omega}h^2$ at a
level of less than 1 per cent, where ${\Omega}$ is the density of the
universe in units of the critical
density and H$_0=100 h~ {\rm km.~ sec}^{-1}{\rm Mpc}^{-1}$. The same is
true for $m_{\nu}$ in
the decaying neutrino theory. Accordingly in this discussion we shall work
to four significant
figures where necessary.

\section{Updated Input Quantities}

\vskip 0.3 cm
\indent
\item {\bf (i)} {$n_{\nu}/n_{\gamma}$}
\vskip 8.25pt

The standard value of this quantity is strongly influenced by the fact that
in the early universe $e^-
- e^+$ pairs annihilated permanently only after neutrinos had already
decoupled from the
primordial heat bath. The resulting annihilation photons (or equivalently
their entropy) then boosted
the photons in the heat bath but not the decoupled neutrinos. One then
finds (Alpher et al 1953) that
$$
n_{\nu}= {3\over 11}n_{\gamma}$$

\noindent
for each of the three neutrino types ${\nu}_e, {\nu}_{\mu}$ and
${\nu}_{\tau}$.The decaying
neutrino must be either ${\nu}_{\mu}$ or ${\nu}_{\tau}$, since
$m_{{\nu}_e}$ is known to be too low
for decay photons from ${\nu}_e$ to be able to ionise hydrogen.
We now need to know how accurate is the factor $3/11$. This question has
been answered by a
number of recent calculations which use the Boltzmann equation to take into
account the
incompleteness of the decoupling of the neutrinos from the heat bath when
the electron pairs
annihilated permanently. We shall use the detailed calculations of
Hannestad and Madsen (1995)
who found that, for ${\nu}_{\mu}$ or ${\nu}_{\tau},~n_{\nu}$  is increased
by $0.25$ percent.
We will take this correction into account, and will assume that any further
correction is negligible
for our purposes.

\vskip 8.25pt
\item {\bf (ii)} {$n_{\gamma}$}
\vskip 8.25pt

We know from COBE that the cosmic microwave background has an accurately
thermal spectrum;
observed RMS deviations are less than 50 parts per million of the peak
brightness (Fixsen et al
1996). For such a spectrum of temperature T
$$
n_{\gamma}=16{\pi}{\zeta}\left(3\right)\left(kT/hc\right)^3\;,
\eqno(1)$$
where ${\zeta}\left(3\right)=1. 202\;.$ The most accurate value for  T has
been given by Fixsen et al
(1996), namely
$$
T=2.728\pm0.004 K\;.
\eqno(2)$$

Thus the present uncertainty in T is only about $1/7$ per cent. We
therefore use (1) to calculate
$n_{\gamma}$ with T given by (2), so that $n_{\gamma}$ has an uncertainty
of about 3/7 per
cent. We find that
$$n_{\gamma}=412\pm 2 {\rm cm}^{-3}\;.$$
Hence
$${3\over 11}n_{\gamma}=112. 4\pm 0. 5 {\rm cm}^{-3}\;.$$
When we correct for the partial coupling of the neutrinos we obtain
$$n_{\nu}=112.6\pm 0. 5 {\rm cm}^{-3}\;.$$

\vskip 8.25pt
\item {\bf (iii)} {$n_b$}
\vskip 8.25pt

We use a value for $n_b$ derived from recent measurements of the abundances
of D,  He$^4$ and
Li$^7$, and the theory of big bang nucleosynthesis. For D we refer to Hata
et al (1997) and
Songaila et al (1997) who discuss two competing derived values of
${\Omega}_b h^2$: a low
value of $\left(7.5\pm 2. 5\right)\times10^{-3}$ and a high value of
$\left(25\pm 5\right)\times10^{-3}$. However, as Hata et al point out, the
high value is inconsistent
with the value of ${\Omega}_b h^2$ derived from recent measured abundances
of He$^4$ and
Li$^7,$ and also with the known number of neutrino types. We therefore adopt
the low value of
${\Omega}_b h^2=\left(7. 5\pm 2.5\right)\times10^{-3}$ from the deuterium
measurements.
For the abundance of He$^4$ we refer to Izotov et al (1997) and Olive et al
(1996). From their
discussion we adopt a derived value for ${\Omega}_{b}h^2$ of $\left(7\pm
1\right)\times10^{-
3}$ from the He$^4$ measurements.
Finally we consider the abundance of Li$^7$. For a given observed abundance
the theory yields
two alternative possible values of ${\Omega}_{b}h^2\;.$ Following the
recent work of Bonifacio
and Molaro (1997) we take for these two alternatives $6.2^{+1.8}_{-1.1}
\times10^{-3}$
and $14.6^{+2.9}_{-3.3}\times10^{-3}$. In order to be consistent with the D and He$^4$
values of ${\Omega}_{b}h^2$ we adopt the value
$6.2^{+1.8}_{-1.1}\times10^{-3}$.
When we combine the results derived from the abundances of the three
isotopes, we find
consistency if we adopt
$${\Omega}_{b}h^2=\left(7\pm 1\right)\times10^{-3}\;.$$

Compared to our final value for ${\Omega}h^2$ of $0. 3$ (see below) we note
that the uncertainty in
our adopted value for ${\Omega}_{b}h^2$ contributes a relative uncertainty
of only $0. 3$ per
cent to the final result.

\vskip 8.25pt
\item {\bf (iv)}{$m_{\nu}$}
\vskip 8.25pt

We first relate $m_{\nu}~$to ${\Omega}_{\nu}h^2$. From general relativity we
have that
${{8{\pi}}\over {3}}G~\rho_{crit}={\rm H}_0^2$, and so
$$\rho_{crit}=1. 880\times10^{-29}h^2 {\rm gm~ cm}^{-3}\;.$$
We also have from (ii) that
$${\rho}_{\nu}=\left(112.6\pm 0.5 \right) m_{\nu} ~{\rm mass~ units~
cm}^{-3}\;.$$

Since a mass unit of  1 eV corresponds to $1. 783\times10^{-33}$ gm,  we
find that
$$m_{\nu}=\left(93. 6\pm 0. 4\right) {\Omega}_{\nu}h^{2}{\rm ev}.$$

To determine $m_{\nu}$ we follow Sciama (1990b) and use the observed upper
limit on the
extragalactic hydrogen-ionising flux F at zero red shift to derive an upper
limit on the energy
$E_{\gamma}$ of a decay photon in the rest frame of the decaying neutrino.
We then assume that
the mass $m_2$ of the residual neutrino in the decay is much less than
$m_{\nu}$ so that we can
write $m_{\nu}=2E_{\gamma}$.  To keep the uncertainty resulting from this
step well below 1 per
cent, we explicitly assume that $m_{\nu}/m_{2}>30$ (as would be easily
satisfied, for example,
in the see-saw model for neutrino masses (Yanagida 1978, Gell-Maun et al 1979)).
The main change from Sciama (1990b) results from the recent establishment
of a six times more
stringent observational upper limit on F than we used there.
Following Vogel et al (1995) we now adopt
$$F<10^5 {\rm photons ~cm}^{-2}{\rm sec}^{-1}\;.$$

The contribution $F_{\gamma}$ of cosmological decay photons to F is
governed by the excess of
$E_{\gamma}$ over $13.6 ~{\rm ev}$, because the red shift eventually
reduces the energy of a decay photon to below the
ionisation potential of hydrogen. Writing
$$E_{\gamma}=13. 6+\epsilon~ {\rm eV}\; ,$$
we have
$$
F_{\gamma}={n_\nu\over\tau}{c\over{H_0}}{\epsilon\over{13.6}},
$$
where ${\tau}$ is the decay lifetime, which in our theory satisfies
${\tau}<3\times10^{23}{\rm
sec}$ (Sciama 1993). We thus require that
$${\epsilon}<0. 39  {\rm h ~eV}.$$
Since ${\epsilon}$ contributes a relatively small fraction to $E_{\gamma}$,
we here insert our
final value for h of $0. 55$ (see below) to obtain
$${\epsilon}<0. 21 ~{\rm eV}\;.$$
Hence we can write
$$E_{\gamma}=13.7\pm 0.1~{\rm eV}\;,$$
and
$$m_{\nu}=27. 4\pm 0.2 ~{\rm eV}\;,$$
so that
$${\Omega}_{\nu}h^2=0.293{\pm 0.003}\;.$$

\vskip 8.25pt
\item {\bf (v)} {${\Omega}h^2$}
\vskip 8.25pt

To obtain ${\Omega}h^2$ we combine ${\Omega}_{\nu}h^2$ and
${\Omega}_{b}h^2$ to obtain
$${\Omega}h^2=0. 300\pm 0. 003\;.$$
Thus the density parameter of the universe ${\Omega}h^2$ is determined with a
precision of 1 per cent in the decaying neutrino theory.
\vskip 0.8 cm
\noindent
{\bf 3 {\hskip 0.4 cm}THE DERIVATION OF H$_{0}$ AND $t_{0}$}
\vskip 0.3 cm
To derive H$_{0}$ and $t_{0}$ from our value for ${\Omega}h^2$ we
proceed as
follows. First we assume from the observations that
$$\left(a\right)h\geq 0. 5$$
and
$$\left(b\right)t_0\geq 12 ~{\rm Gyr}$$
(e.g. Chaboyer et al 1996). From (a) and ${\Omega}h^2=0.3$ we derive that
${\Omega}\leq 1.2$ and so (for a zero
cosmological constant $\lambda$) that $t_0\leq 13.4$ Gyr.
From (b) and ${\lambda}=0$ we
derive that ${\Omega}\geq 1$ and $h\leq 0.55$. Combining these results we obtain
$${\rm H}_0=52.5\pm 2. 5 ~{\rm km.~ sec}^{-1}{\rm Mpc}^{-1}\;,$$
$$t_0=12. 7\pm 0. 7 ~{\rm Gyr}\;,$$
and
$${\Omega}=1. 1\pm 0. 1\;.$$
Since ${\Omega}$ is required to be so close to 1 it is tempting to assume
that it is exactly 1 (or at
least 1 to within the level of uncertainty to which we are working here).
In that case the
uncertainties in H$_0$ and $t_0$  are controlled by our one per cent
uncertainty in
${\Omega}h^2$, and so we find that for ${\lambda}=0$
$${\rm H}_0=54. 8\pm 0. 3~ {\rm km.~ sec}^{-1}{\rm Mpc}^{-1}\;,$$
and
$$t_0=11. 96\pm 0. 06 ~{\rm Gyr}\;,$$
(where 1 year is defined to be $3. 15\times10^7$ seconds).
It is gratifying that these precise predictions for the values of H$_0,
t_0$ and ${\Omega}$ from
the decaying neutrino theory are compatible with current observational
estimates of these quantities.

\section*{Acknowledgments}

I am grateful to M.U.R.S.T. for their financial support of this work.

\bye